\documentstyle[aps,twocolumn, floats]{revtex}
\newcommand{\be}{\begin{equation}}
\newcommand{\ee}{\end{equation}}
\input psfig.sty
\begin{document}
\draft
\title{Bond operator mean field theory of the half filled Kondo lattice model}
\author{Christoph Jurecka, Wolfram Brenig}

\address{Institut f\"ur Theoretische Physik,
Technische Universit\"at Braunschweig, 38106 Braunschweig, Germany}

\date{\today}
\maketitle
\begin{abstract}  
We present a bond-operator mean field theory for the Kondo lattice model
at half filling in two (2D) and three (3D) dimensions. A continuous quantum
phase transition from an antiferromagnetic to a spin-gapped singlet ground
state is found at $J/t=1.505$ ($1.833$) in $2D$ ($3D$). Additionally we
evaluate the quasiparticle dispersions as well as the staggered magnetic
moment and provide a comparison with complementary numerical approaches.
\end{abstract}


The Kondo Lattice Model (KLM) describes the exchange-scattering of a band of
itinerant conduction-electrons at a lattice of localized magnetic moments.  It
serves as a basic model for heavy fermion materials in the integral-valent
limit \cite{Fulde88}. At half filling of the conduction band it is believed to
give a description of Kondo insulators, which have been of considerable
interest in recent years\cite{Aeppli92}. In one dimension the ground state at
half filling is a spin singlet for all values of exchange-scattering and
conduction-band widths\cite{tsunetsugu97}. In higher dimensions  it has been
suggested early on, that the competition between Kondo screening and the
Ruderman-Kittel-Kasuya-Yoshida (RKKY) interaction leads to a quantum phase
transition between a global spin-singlet and an antiferromagnetically ordered
phase\cite{doniach77,lacroix79}.  This scenario has been corroborated  in two
dimensions by variational\cite{wang94} calculations, series
expansion\cite{shi95} and mean-field\cite{zhang00-2} approaches. Numerically
exact results in 2D have been obtained recently by
QMC\cite{assaad99,capponi00}.  In $3D$ only series expansion is
available\cite{shi95}.

The purpose of this work is to introduce a novel mean-field theory for the KLM
at half filling in two and three dimensions. In contrast to other mean-field
calculations\cite{lacroix79,zhang00-2} our approach is based on a bond-operator
representation of the KLM which is suitable for strong exchange scattering and
has proven to be useful in dimerized spin systems \cite{sachdev90}. Moreover,
our treatment goes beyond recent mean-field work focusing on the Kondo-necklace
problem which neglects conduction-electron charge fluctuations
\cite{Zhang00}. The KLM reads
\be
H=-t \sum_{\{i,j\}, \sigma} c^\dagger_{i,\sigma}c^{\phantom\dagger}_{j,\sigma}+
J\sum_i {\bf S}_{i,c}{\bf S}_{i,f}
\label{e1} 
\ee
with spin operators ${\bf
S}_{c(f)}=\frac{1}{2}\sum_{\sigma,\sigma'}c(f)^\dagger_\sigma
\tau_{\sigma,\sigma'} c(f)^{\phantom\dagger}_{\sigma'}$ 
and destruction(creation) operators $c^{(\dagger)}_{i,\sigma}$ and
$f^{(\dagger)}_{i,\sigma}$ for itinerant $c$ and localized $f$ electrons of
spin $\sigma$ at sites $i$. An f-occupation of exactly one per lattice site is
implied.

The local Hilbert space consists of one $f$ electron with spin up or down and
additionally up to two itinerant electrons. The resulting eight possible states
can be created by applying the following operators onto the vacuum $|0\rangle$
of an empty site
\begin{eqnarray}s^\dagger|0\rangle&=&\frac{1}{\sqrt{2}}(
c^\dagger_\uparrow f^\dagger_\downarrow+f^\dagger_\uparrow 
c^\dagger_\downarrow)|0\rangle \nonumber\\
t^\dagger_x|0\rangle&=&\frac{-1}{\sqrt{2}}(c^\dagger_\uparrow 
f^\dagger_\uparrow-c^\dagger_\downarrow f^\dagger_\downarrow)
|0\rangle \nonumber\\
t^\dagger_y|0\rangle&=&\frac{i}{\sqrt{2}}(c^\dagger_\uparrow 
f^\dagger_\uparrow+c^\dagger_\downarrow f^\dagger_\downarrow)
|0\rangle \nonumber\\
t^\dagger_z|0\rangle&=&\frac{1}{\sqrt{2}}(c^\dagger_\uparrow 
f^\dagger_\downarrow+c^\dagger_\downarrow f^\dagger_\uparrow)
|0\rangle \nonumber\\
a^\dagger_\sigma|0\rangle&=&f^\dagger_\sigma|0\rangle \nonumber\\
b^\dagger_\sigma|0\rangle&=&c^\dagger_\uparrow c^\dagger_\downarrow 
f^\dagger_\sigma|0\rangle.
\end{eqnarray}
where the operators $s$ and $t$ are equivalent to the so-called bond operators
of\cite{sachdev90} and are assumed to obey {\em bosonic} commutation relations.
The {\em fermionic} operators $a$ and $b$ have been introduced first
in\cite{eder97} and label states with one or three electrons per site. In order
to suppress unphysical states a constraint of no double occupancy
\be
s^\dagger_j s^{\phantom\dagger}_j+\sum_\alpha t^\dagger_{\alpha,j} 
t^{\phantom\dagger}_{\alpha,j}+\sum_\sigma a^\dagger_{\sigma,j}
a^{\phantom\dagger}_{\sigma,j}+\sum_\sigma b^\dagger_{\sigma,j}
b^{\phantom\dagger}_{\sigma,j}=1
\label{e11}
\ee
has to be fulfilled. The original fermion and spin operators are represented by
\begin{eqnarray}
c^\dagger_{j,\sigma}=p_\sigma\frac{1}{\sqrt{2}}[(s^\dagger_j+p_\sigma 
t^\dagger_{z,j})a^{\phantom\dagger}_{-\sigma,j}-(t^\dagger_{x,j}+ p_\sigma i 
t^\dagger_{y,j})a_{\sigma,j}]\nonumber \\
-\frac{1}{\sqrt{2}}[b^\dagger_{\sigma,j}(s^{\phantom\dagger}_j-p_\sigma 
t^{\phantom\dagger}_{z,j})-b^\dagger_{-\sigma,j}(t_{x, j}^{\phantom\dagger}+
p_\sigma it_{y,j}^{\phantom\dagger})] \label{e8}
\end{eqnarray}
\begin{eqnarray}
S^c_{\alpha,j}&=&\frac{1}{2}(-t^\dagger_{\alpha, j} 
s^{\phantom\dagger}_j-s^\dagger_j t^{\phantom\dagger}_{\alpha, j}-i 
\sum_{\beta, \gamma} \epsilon_{\alpha\beta\gamma} t^{\dagger}_{\beta, j} 
t^{\phantom\dagger}_{\alpha, j}) \label{e9}\\
S^f_{\alpha,j}&=&\frac{1}{2}(t^\dagger_{\alpha, j} 
s^{\phantom\dagger}_j+s^\dagger_j t^{\phantom\dagger}_j-i  
\sum_{\beta, \gamma}\epsilon_{\alpha\beta\gamma} t^{\dagger}_{\beta, j} 
t^{\phantom\dagger}_{\alpha, j})\nonumber \\&+&S^a_{\alpha, j}+S^b_{\alpha, j} 
\label{e10}
\end{eqnarray}
with ${\bf S}_{a(b),j}=\frac{1}{2}\sum_{\sigma,\sigma'}a(b)^\dagger_{\sigma,j}
\tau_{\sigma,\sigma'} a(b)^{\phantom\dagger}_{\sigma',j}$ and $p_\uparrow=1$,
$p_\downarrow=-1$. This mapping is exact and preserves the proper
commutation relations provided that $s$ and $t$ are bosons, $a$ and $b$
are fermions, and (\ref{e11}) is satisfied.

Rewriting the KLM in terms of (\ref{e11},\ref{e8}-\ref{e10}) leads to a
strongly correlated boson-fermion model which cannot be solved exactly, but
approximations have to be made. Analogous models in the context of
quantum-disordered ground states in dimerized spin systems have recently been
treated by various schemes of approximation
\cite{gopalan94,jurecka00,eder98,kotov98}. Here we are interested in a
mean-field description of the transition from an antiferromagnetic state to a
spin-singlet regime. The latter can be described by allowing for a condensate
of singlets\cite{sachdev90}
\be
\langle s^{\phantom\dagger}_j \rangle=\langle s^\dagger_j \rangle=s
\label{e12}
\ee
while the antiferromagnetically ordered phase requires an additional 
condensation of one of the triplets \cite{normand97}
\be
\langle t^{\phantom\dagger}_{z, j} \rangle=\langle t^\dagger_{z,j} 
\rangle=m_j=(-1)^j m
\label{e13}
\ee
For the remainder of this work we assume a bipartite lattice structure with the
factor of $(-1)^j$ being a shorthand for '$+1(-1)$' on the A(B) sublattice.

Inserting (\ref{e12},\ref{e13}) into (\ref{e8}-\ref{e10}) and (\ref{e1}) we
drop all longitudinal and transverse spin-fluctuations, i.e., we only keep the
mean-field values of $s$ and $t$ in addition to the fermions $a$ and $b$. We
obtain
\begin{eqnarray}
H&=&-\frac{t}{2}\sum_{\{i,j\},\sigma}(-sp_\sigma+m_i)(-sp_\sigma+m_{j})\times
\nonumber\\ 
&&\qquad\times(a^{\phantom\dagger}_{\sigma,i} a^\dagger_{\sigma,j}+
b^\dagger_{\sigma,i}b^{\phantom\dagger}_{\sigma,j})\nonumber+h.c.\\
&&-\frac{t}{2}\sum_{\{i,j\},\sigma}(-sp_\sigma+m_i)(sp_\sigma+m_{j})
\times\nonumber\\&&\qquad\times(-p_\sigma a^{\phantom\dagger}_{\sigma,i} 
b^{\phantom\dagger}_{-\sigma,j}+p_\sigma b^\dagger_{\sigma,i}
a^{\dagger}_{-\sigma,j})+h.c.\nonumber\\&&
-\frac{3}{4}JNs^2+\frac{1}{4}JNm_i^2\nonumber\\&&
+\sum_{i, \sigma} \mu_i (s^2+m_i^2+a^\dagger_{\sigma,i}
a^{\phantom\dagger}_{i,\sigma}+b^\dagger_{\sigma,i}
b^{\phantom\dagger}_{\sigma,i}-1)\nonumber\\&&
+\lambda \sum_{i, \sigma} (b^\dagger_{\sigma,i} b^{\phantom\dagger}_{\sigma,i}-
a^\dagger_{\sigma,i}a^{\phantom\dagger}_{\sigma,i})
\label{e14}
\end{eqnarray}
The first sum $\propto t$ describes hopping of particle- or hole-like
fluctuations with a dispersion renormalized by a factor of $(s^2-m^2)$, taking
into account the alternation of the sign of $m_i$. This prefactor leads to a
reduction of the fermionic kinetic energy in the antiferromagnetic phase. The
second sum $\propto t$ generates the mean-field analog of the intermediate
state of the RKKY process and its hermitian conjugate, i.e., the
destruction(creation) of two adjacent two-particle states accompanied by the
creation(destruction) of a pair of two adjacent three- and one-electron states.
Second order processes from this term drive the transition into the magnetic
state. Thus we believe that this mean-field Hamiltonian incorporates the basic
ingredients to induce the quantum-phase transition of the KLM.  Furthermore in
(\ref{e14}) we have introduced the usual chemical potential $\lambda$ to set
the global particle density and a local Lagrange multiplier $\mu_i$ in order to
to enforce the constraint (\ref{e11}).

To diagonalize the Hamiltonian we first replace the local Lagrange multiplier
by a global one $\mu_i\rightarrow\mu$.  Second we switch to an
antiferromagnetic unit cell and finally perform a generalized Bogoliubov
transformation to eliminate particle-number non-conserving term of type
$ab(a^\dagger b^\dagger)$.  This leads to $4$ bands $\omega_{1,2}({\bf
k})=\lambda\pm E_1({\bf k})$ and $\omega_{3,4}({\bf k})=\lambda\pm E_2({\bf
k})$ which are twofold degenerate by spin-z quantum number
\begin{eqnarray}
E_{1,{\bf k}}&=&\sqrt{\mu^2+\frac{1}{2}\epsilon_{\bf k}^2(m^2+s^2)^2-
2W_{\bf k}} \nonumber \\
E_{2,{\bf k}}&=&\sqrt{\mu^2+\frac{1}{2}\epsilon_{\bf k}^2(m^2+s^2)^2+
2W_{\bf k}}\nonumber\\
W_{\bf k}&=&\sqrt{\frac{1}{4}\mu^2(m^2-s^2)^2\epsilon_{\bf k}^2+
\frac{1}{16}\epsilon_{\bf k}^4(m^2+s^2)^4}
\label{e16}
\end{eqnarray}
Here $\epsilon_{\bf k}=-2t\sum_{d=1}^D \cos k_d$. At half filling the lower two
bands, i.e. $\omega_{2,4}$, are completely filled while the upper two bands ,
i.e. $\omega_{1,3}$, are empty.  This leads to a ground state energy of
\begin{eqnarray}
\frac{E}{N}&=&-\frac{3}{4}Js^2+\frac{1}{4}Jm^2+\mu (s^2+m^2+1) \nonumber \\
&&-\frac{1}{2N}\sum_{\bf k} 2E_{1, {\bf k}}-\frac{1}{2N}\sum_{\bf k} 
2E_{2, {\bf k}}
\end{eqnarray}
which is independent of $\lambda$.

\begin{figure}[t]
\centerline{\psfig{file=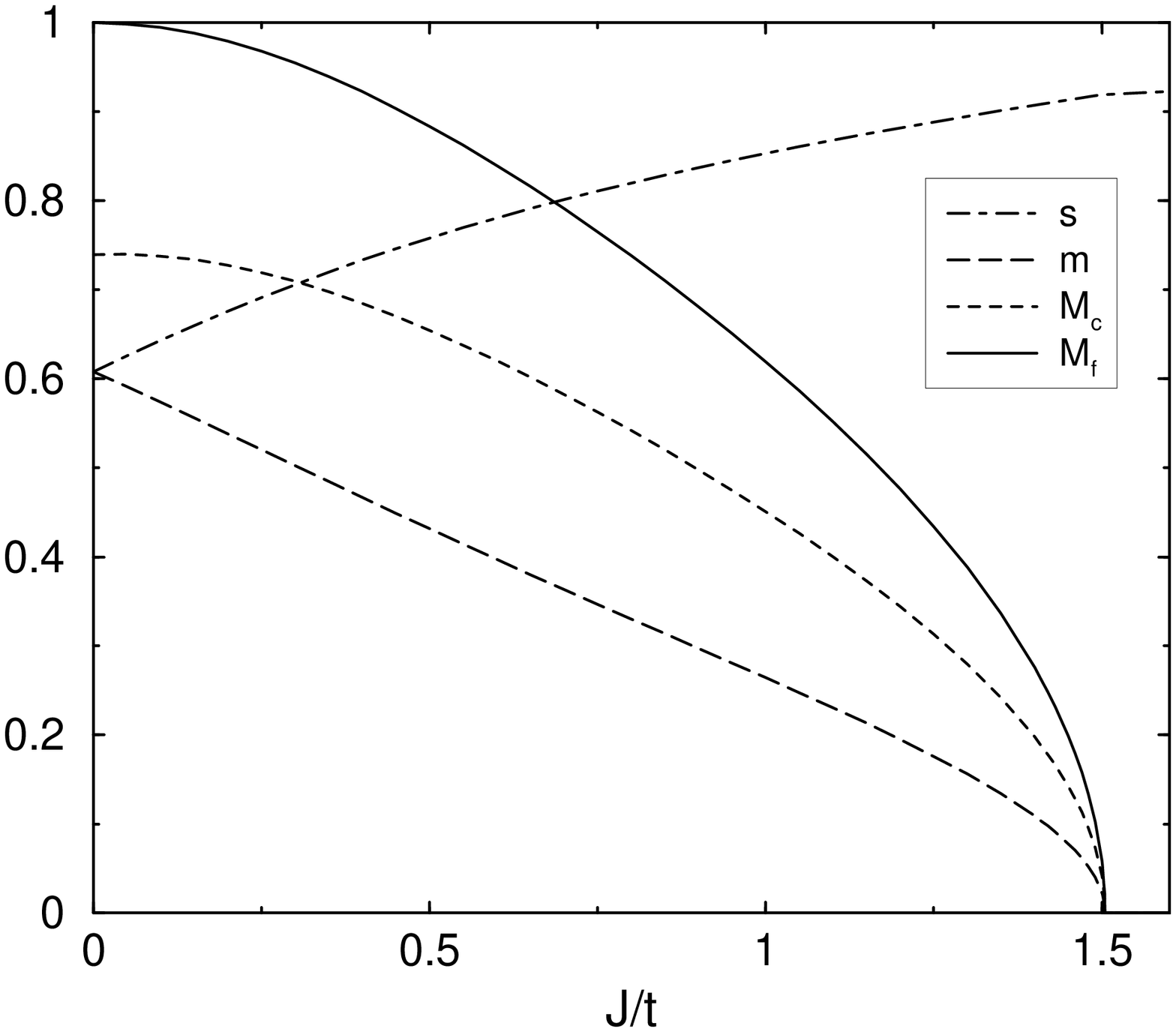,width=0.9\columnwidth,angle=-90}}
\caption{Mean-field order parameter $s$, $m$ and the staggered moments
$M_c$ and $M_f$ as function of $J/t$ for the $2D$ KLM.}
\label{fig1}
\end{figure}
 
The staggered magnetization $M_{c(f)}$ of the $c(f)$ electrons is obtained from
a direct evaluation of the corresponding matrix elements in the mean-field
ground state
\begin{eqnarray}
M_c&=&\frac{2}{N}\sum_n(-1)^n\langle S^c_{z,n}\rangle=2ms\nonumber\\
M_f&=&\frac{2}{N}\sum_n(-1)^n\langle S^f_{z,n}\rangle=\nonumber\\&=&
2ms+\frac{1}{N}\sum_{\bf k}\frac{2 \epsilon_{\bf k}^2\mu m s(s^2+m^2)}
{E_{1,{\bf k}}E_{2,{\bf k}}(E_{1,{\bf k}}+E_{2,{\bf k}})}.
\end{eqnarray}
The second term in $M_f$ is due to a staggering of the spin density of the $a$
and $b$ fermions which develops at non-zero values of $m$ due to the the second
sum $\propto t$ in (\ref{e14}). This terms induces a spin-dependent
non-diagonal momentum-space component in the $a$ and $b$ Greens functions at
the antiferromagnetic nesting wave-vector.

The mean-field equations
\be
\frac{\partial E}{\partial s}=0 \qquad  \frac{\partial E}{\partial m}=0
\qquad \frac{\partial E}{\partial \mu}=0
\label{e18}
\ee
have to be solved self-consistently for the order parameters $s$, $m$ and
$\mu$. In the magnetic phase ($m\neq0$) the system of equations (\ref{e18}) can
be written as
\begin{eqnarray}
0&=&2J+\frac{1}{2N}\sum_{\bf k} \frac{\epsilon_{\bf k}^2 \mu^2(s^2-m^2)}
{W_{\bf k}}\left(\frac{2}{E_{2, {\bf k}}}-\frac{2}{E_{1, {\bf k}}}\right)
\nonumber\\
0&=&-J+4\mu-\frac{1}{2N}\sum_{\bf k} 2\epsilon_{\bf k}^2(m^2+s^2)
\left(\frac{2}{E_{2, {\bf k}}}+\frac{2}{E_{1, {\bf k}}}\right)\nonumber\\
&&-\frac{1}{2N}\sum_{\bf k} \frac{\epsilon_{\bf k}^4(m^2+s^2)^3}
{2W_{\bf k}}\left(\frac{2}{E_{2, {\bf k}}}-\frac{2}{E_{1, {\bf k}}}
\right)\nonumber\\
0&=&s^2+m^2+1-\frac{1}{2N}\sum_{\bf k} \mu \left(\frac{2}{E_{2, {\bf k}}}
+\frac{2}{E_{1, {\bf k}}}\right)\nonumber\\
&&-\frac{1}{2N}\sum_{\bf k} \frac{\epsilon_{\bf k}^2\mu(s^2-m^2)^2}
{4W_{\bf k}}\left(\frac{2}{E_{2, {\bf k}}}-\frac{2}{E_{1, {\bf k}}}\right)\;\;,
\label{e19}
\end{eqnarray}
while for the disordered one ($m=0$) it simplifies considerably 
\begin{eqnarray}
0&=&-\frac{3}{2}J+2\mu-\frac{1}{N}\sum_{\bf k} \frac{2\epsilon_{\bf k}^2s^2}
{\sqrt{4\mu^2+\epsilon_k^2s^4}}\nonumber\\
0&=&s^2+1-\frac{1}{N}\sum_{\bf k}\frac{4\mu}{\sqrt{4\mu^2+
\epsilon_{\bf k}^2s^4}}.
\label{e20}
\end{eqnarray}

For $\epsilon_{\bf k}=0$, i.e. $t=0$, only (\ref{e20}) has a solution.  This
solution also provides for the correct quasiparticle gap, i.e.
$\mu=\frac{3}{4}J$. Upon increasing $t$ charge fluctuations, i.e.  creation of
$a$- and $b$-excitations contribute to the ground state reducing the stability
of the singlet state.  At $J/t\rightarrow 0$ the ground state stems from the
solution of (\ref{e19}) with $s=m$. This state displays a complete polarization
of the $f$ spins, while the $c$ spin density is polarized only partially.  The
latter is consistent with the itinerant character of the $c$ fermions, implying
a finite density of empty and doubly occupied sites, i.e.  a finite density of
$a$ and $b$ fermions.  Since $s=m$ at $J/t=0$ the diagonal part of the kinetic
energy of the $a$ and $b$ particles vanishes at this point.

At intermediate $J/t$ we determine the ground state by solving
(\ref{e19},\ref{e20}) numerically. Fig. \ref{fig1} shows the singlet and
triplet order parameters as well as the staggered magnetizations for the $2D$
KLM.  We observe a continuous quantum phase transition from the singlet to the
magnetic phase at $(J/t)_c=1.505$. This is in good agreement with the data of a
recent QMC study\cite{assaad99} which has determined the phase transition to
occur at $(J/t)_c=1.45\pm0.05$. Similar values of $(J/t)_c$ have also been
reported from variational Monte-Carlo simulations\cite{wang94}
($(J/t)_c=1.4\pm0.1$) and series expansion\cite{shi95}($(J/t)_c=1.43\pm0.2)$.
From Fig. \ref{fig1} it can be seen that the maximum magnetization, i.e. $s=m$,
prevails only at $J=0$ with a continuous increase of $s$ vs. $m$, i.e.
screening of the local moment, to occur as $J$ approaches $J_c$. Such
coexistence of Kondo  screening and antiferromagnetic order has also been
reported recently from QMC calculations\cite{capponi00} for all $J<J_c$ and
within a small window of values of $J$ from a mean-field study\cite{zhang00-2}.

\begin{figure}[t]
\centerline{\psfig{file=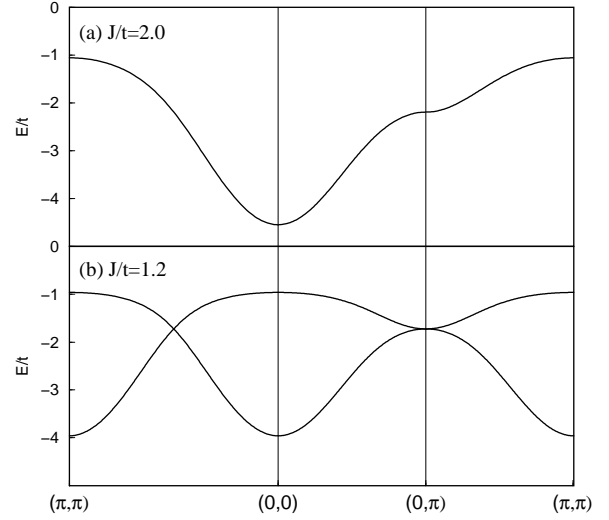,width=0.9\columnwidth,angle=-90}}
\vspace{0.5cm}
\caption{Quasiparticle dispersions for (a) J/t=1.2 and (b) J/t=2.}
\label{fig3}
\end{figure}

Fig. \ref{fig3} shows the quasiparticle dispersion of the occupied bands for
two values of $J/t$ which are in the singlet and the magnetic phase, i.e. (a)
$J/t=2$ and (b) $J/t=1.2$ respectively. These bands are split by a gap from the
unoccupied bands which are located symmetrically reflected along the line
$E/t=0$ at positive energies. For $J>J_c$ the four bands $\omega_{1...4} ({\bf
k})$ collapse onto only two bands by a mere backfolding which has been carried
out in fig. \ref{fig3}(a) leaving a single occupied band to be displayed. For
$J<J_c$ two distinct bands are present throughout the entire magnetic Brillouin
zone. Since the Hamiltonian (\ref{e14}) incorporates scattering with a magnetic
wave-vector ${\bf k} =(\pi,\pi)$ in the off-diagonal terms only, no additional
gap opens along the line ${\bf k}_N$ with $k_{N,x}+k_{N,y}=\pi$, i.e.  $W_{{\bf
k}_N}=0$ in (\ref{e16}). In this context we note, that the interpretation of
the band-gap in this mean-field theory changes {\em quasi-continuously} from a
gap induced by singlet formation at $J>J_c$ to a magnetic gap as
$J\rightarrow0$.

\begin{figure}[t]
\centerline{\psfig{file=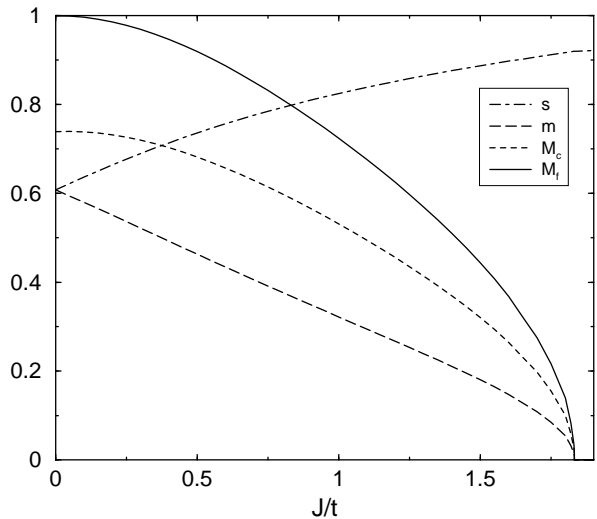,width=0.9\columnwidth,angle=-90}}
\caption{Mean-field order parameter $s$, $m$ and the staggered moments 
$M_{c}$ and $M_{f}$ as function of $J/t$ for the $3D$ KLM.}
\label{fig2}
\end{figure}

In fig. \ref{fig2} we display a set of results identical to that of fig.
\ref{fig1} however for the 3D case. Here the phase transition occurs at 
a slightly larger value of $(J/t)_c=1.833$, which is in reasonable agreement
with an estimate of $(J/t)_c=2.04\pm0.16$ reported from series
expansion\cite{shi95}. Again, the $f$ spins are fully polarized as
$J\rightarrow0$, while the maximum value of $M_c$ is nearly identical to that
of the $2D$ case.

To conclude several comments are in order. First, and very much in contrast to
usual approaches to the KLM \cite{lacroix79,Fulde88} our method is best suited
for the limit of strong and quasi {\em local} Kondo screening at large $J/t$.
In that limit the Kondo effect can be viewed as a molecular singlet formation
within each unit cell resulting in an algebraic energy scale of order $J$,
rather than the usual Kondo energy-scale $T_K \sim t \exp(-t/J)$. While the
large-$J$ limit may obliterate some of the subtleties genuine to the Kondo
effect at $J\ll t$, we believe that it is a superior starting point for
analytic studies of the quantum phase transition in the  2D and 3D KLM since
this transition occurs at $J/t>1$.  Second, we note that while we have
neglected quantum fluctuations of the triplet order parameter, it would be
interesting to incorporate them into future studies. In particular, it is
conceivable that transverse fluctuations due to the $t_{x,y}$-operators will
reduce the staggered magnetization in the ordered phase. This seems consistent
with QMC finding a smaller magnetization\cite{assaad99} than we observe within
the mean-field approach.  Finally an extension of the scheme presented here to
incorporate Coulomb correlations or finite doping, off half filling, into the
conduction band are open issues.

In summary we have studied the KLM using a novel bond-operator mean-field
theory. In good agreement with complementary approaches we find a quantum phase
transition at $(J/t)_c =1.505(1.833)$ in 2(3) dimensions.  In addition we have
evaluated the magnetization in the ordered phase and the quasiparticle
dispersions.

This research was supported in part by the Deutsche Forschungsgemeinschaft
under Grant No. BR 1084/1-1 and BR 1084/1-2.

\end{document}